\documentclass[a4paper,fleqn,usenatbib]{mnras}

\usepackage{newtxtext,newtxmath}

\usepackage[T1]{fontenc}
\usepackage{ae,aecompl}

\usepackage{graphicx}	
\usepackage{amsmath}	
\usepackage{amssymb}	
\usepackage{subfigure}
\usepackage{color}

\title[Observations in models without FLRW backgrounds]{Observations in statistically homogeneous, locally inhomogeneous cosmological toy-models without FLRW backgrounds}

\author[S. M. Koksbang]{
	S. M. Koksbang\thanks{E-mail: koksbang@cp3.sdu.dk}
	\thanks{New affiliation: CP$^3$-Origins, University of Southern Denmark, Campusvej 55, DK-5230 Odense M, Denmark}
	\\
	Department of Physics, University of Helsinki and Helsinki Institute of Physics, P.O. Box 64, FIN-00014 University of Helsinki, Finland\\
}

\date{Accepted XXX. Received YYY; in original form ZZZ}

\pubyear{2020}

\begin{document}
\label{firstpage}
\pagerange{\pageref{firstpage}--\pageref{lastpage}}
\maketitle

\begin{abstract}
Observations are studied in toy-models constituting  exact cosmological solutions to the Einstein equation which are statistically homogeneous but locally inhomogeneous, without an {\em a priori} introduced FLRW background and with ``structures'' evolving fairly slowly. The mean redshift-distance relation and redshift drift along 500 light rays in each of two models are compared to relations based on spatial averages. The relations based on spatial averages give a good reproduction of the mean redshift-distance relation, although most convincingly in the model where the kinematical backreaction is sub-percent. In both models, the mean redshift drift clearly differs from the drift of the mean redshift. This indicates that redshift drift could be an important tool for testing the backreaction conjecture as redshift drift appears to distinguish between local and global effects. The method presented for computing the redshift drift is straightforward to generalize and can thus be utilized to fairly easily compute this quantity in a general spacetime.
\end{abstract}

\begin{keywords}
cosmology: observations -- (cosmology:) large-scale structure of Universe -- cosmology: theory
\end{keywords}



\section{Introduction}
Modern cosmology is based on the Friedmann-Lemaitre-Roberston-Walker (FLRW) solutions to the Einstein equation. The dynamics of the FLRW universe is given by the Friedmann equations (subscripted commas followed by one or more coordinates or indices indicate partial derivatives, and $c=1$)
\begin{equation}
\left( \frac{a_{,t}}{a}\right) ^2 = \frac{8\pi G}{3}\rho - \frac{\kappa }{R_0^2a^2}+\frac{\Lambda}{3}
\end{equation}
\begin{equation}
\frac{a_{,tt}}{a} = -\frac{4\pi G}{3}\left(\rho + 3p \right) +\frac{\Lambda}{3},
\end{equation}
where $\rho$ is the density, $p$ the pressure and $a$ is the scale factor appearing in the FLRW line element, $ds^2 = -dt^2 + a^2\left( \frac{dr^2}{1-k} + r^2d\Omega^2\right)$,
with $k = \kappa r^2/R_0^2$ the curvature parameter ($\kappa =\pm1,0$).
\newline\newline
{\em The inhomogeneous universe:} Unlike the FLRW universes which are spatially exactly homogeneous and isotropic, the real universe is at most spatially {\em statistically} homogeneous and isotropic. This difference is a potentially vital detail because the spatially averaged expansion of a generic inhomogeneous universe in general deviates from FLRW evolution. This deviation is known as cosmic backreaction \citet{bc_review1,bc_review2,bc_review3} and is due to the fact that in general relativity, spatial averages and time derivatives do not commute.
\newline\indent
In the case where averages are computed on hypersurfaces orthogonal to the fluid velocity field and with metric lapse function equal to 1, the equations describing the average evolution of the Universe are (the Buchert equations, \citet{fluid2II})
\begin{align}\label{eq:Friedmann}
\begin{split}
\frac{1}{3}\left\langle \theta\right\rangle^2& = 3\left( \frac{ a_{D,t}}{a_D}\right) ^2= 8\pi G_N\left\langle \rho\right\rangle - \frac{1}{2}\left\langle ^{(3)}R\right\rangle +\Lambda - \frac{1}{2}Q
\\
&\iff \Omega_Q+\Omega_R+\Omega_{\rho}+\Omega_{\Lambda} = 1
\end{split}
\end{align}
\begin{equation}\label{eq:acc}
3\frac{a_{D,tt}}{a_D} = -4\pi G\left\langle \rho+3 p\right\rangle + \Lambda + Q .
\end{equation}
Triangular brackets denote spatial averaging of a scalar, i.e. $\left\langle s\right\rangle :=\frac{\int_D sg^{(3)}d^3x}{\int_Dg^{(3)}d^3x}$, where $g^{(3)}d^3x$ is the infinitesimal spatial volume element, $D$ is the spatial domain of averaging and $s$ is some scalar. The volume averaged scale factor is defined through the proper volume of a spatial averaging domain such that $a_D:=\left( \frac{\int_Dg^{(3)}d^3x}{\int_{D_0}g_0^{(3)}d^3x}\right)^{1/3} $ (subscripted zeros indicate evaluation at present time). As eq. \ref{eq:Friedmann} shows, the evolution of $a_D$ is determined by the spatially averaged local expansion rate, $\left\langle \theta \right\rangle $. Density parameters, $\Omega_x$, are defined by dividing the respective terms in equation \ref{eq:Friedmann} by $3H_D^2:=3\left( \frac{ a_{D,t}}{a_D}\right) ^2$. Note that $\Lambda$ was included in the above equations for completeness but will be set to zero in the studied models.
\newline\indent
By comparing with the Friedmann equations, one sees that there is an extra term in the Buchert equations, namely $Q$ which is known as the kinematical backreaction. The kinematical backreaction is defined by $Q:=\frac{2}{3}\left(\left\langle \theta ^2\right\rangle-\left\langle \theta\right\rangle^2 \right) -2\left\langle \sigma^2\right\rangle $, where $\sigma^2:=\frac{1}{2}\sigma_{\mu\nu}\sigma^{\mu\nu}$ is the shear scalar of the fluid. In addition to the kinematical backreaction, the Buchert equations differ from the Friedmann equations by the curvature term which in the Buchert equations is given by the spatial average of the hypersurface Ricci scalar, $\left\langle ^{(3)}R\right\rangle $, which may evolve differently than the curvature in the Friedmann equations, i.e. differently than simply $\propto a_D^{-2}$.
\newline\newline
It is unknown how cosmic backreaction affects the large scale/average evolution of the Universe: It may turn out to be negligible, but as the equations above show, backreaction can e.g. lead to average accelerated expansion and it has been suggested that backreaction may be the true explanation for the apparent late time accelerated expansion of the Universe. However, a realistic quantification of backreaction is highly non-trivial as it requires a realistic, {\em general relativistic} description of the Universe which is not currently available. Another route to quantifying the importance of backreaction in our universe is through observations. Specifically, several relations have been identified which can test the FLRW assumption observationally \citet{FLRWtest1,FLRWtest2,FLRWtest3}. If observations fail these tests, the Universe cannot be described by an FLRW metric on large scales and backreaction is likely to be important. If observations fulfill these tests, it is however not guaranteed that the Universe is FLRW on large scales, so there is currently no known way of unambiguously falsifying the idea that backreaction has an important impact on the dynamics of the Universe. A main obstacle is that it is not known how to relate spatially averaged quantities to observations. Several methods have been proposed in the literature (e.g. \citet{template1,template2,template3,av_obs1,av_obs2}, see e.g. also the discussions in \citet{koksbang1,koksbang2}), but in order to determine if any of these are accurate, they must be tested using exact solutions to the Einstein equation which are not {\em a priori} based on FLRW backgrounds. The usual Swiss-cheese construction and e.g. relativistic codes based on weak-field approximations do not fulfill this requirement as they are based on pre-specified FLRW backgrounds. In this Letter, exact, inhomogeneous, statistically homogeneous cosmological models are constructed without introducing an FLRW ``background'' and the mean redshift-distance relation and redshift drift along 500 light rays in each model are computed and compared with relations based on spatially averaged quantities.

\section{Model construction}
\begin{figure}
	\centering
	\includegraphics[scale = 0.37]{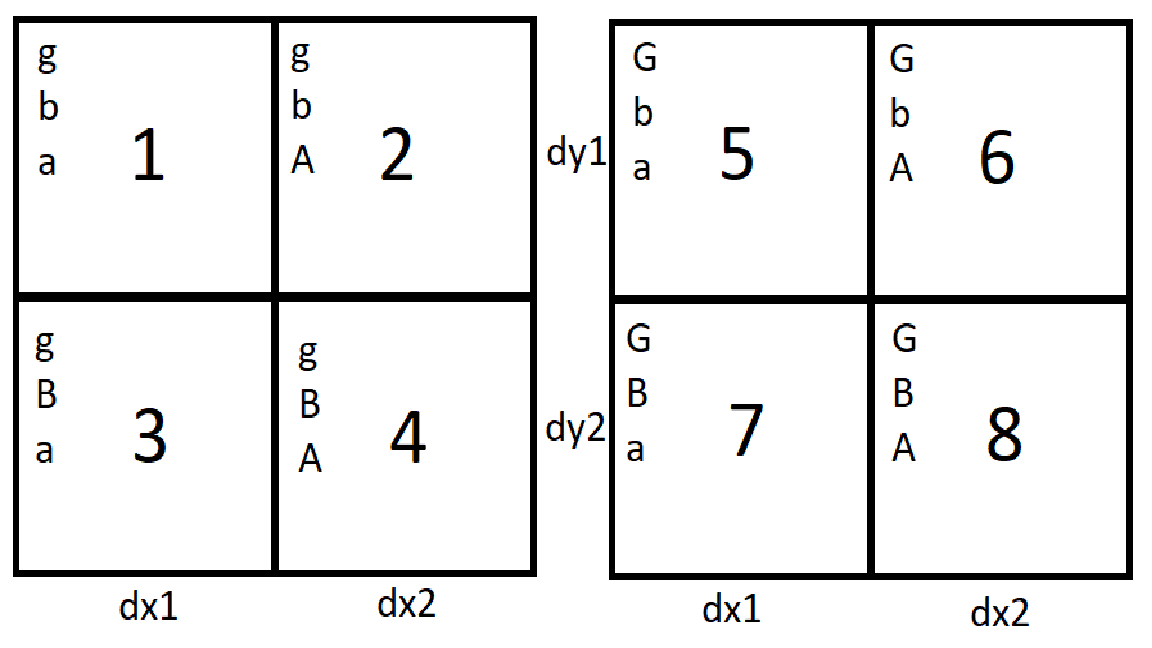}
	\caption{2D rendering of fundamental Bianchi I block with their values of $\alpha, \beta$ and $\gamma$ as well as their size and numbering ($\alpha$ takes the value a or A, $\beta$ b or B, and $\gamma$ g or G). The cubes in the left side of the figure have a comoving height of dz1, and those to the left dz2. The two sets of cubes are stacked on top of each other such that cube 1 is below cube 5 etc.}
	\label{fig:model}
\end{figure}

\begin{figure*}
	\centering
	\subfigure{
		\includegraphics[scale = 0.5]{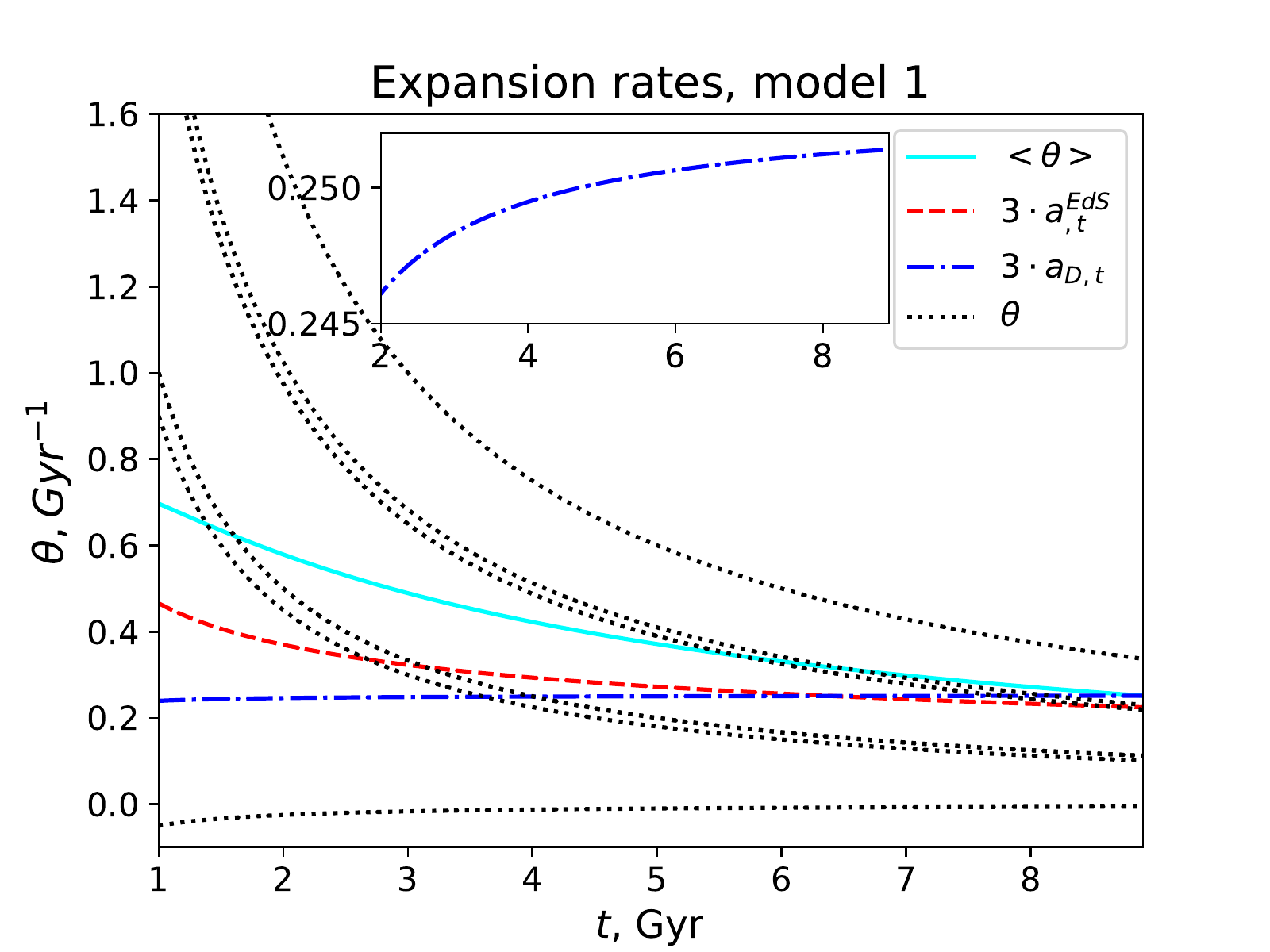}
	}
	\subfigure[]{
		\includegraphics[scale = 0.5]{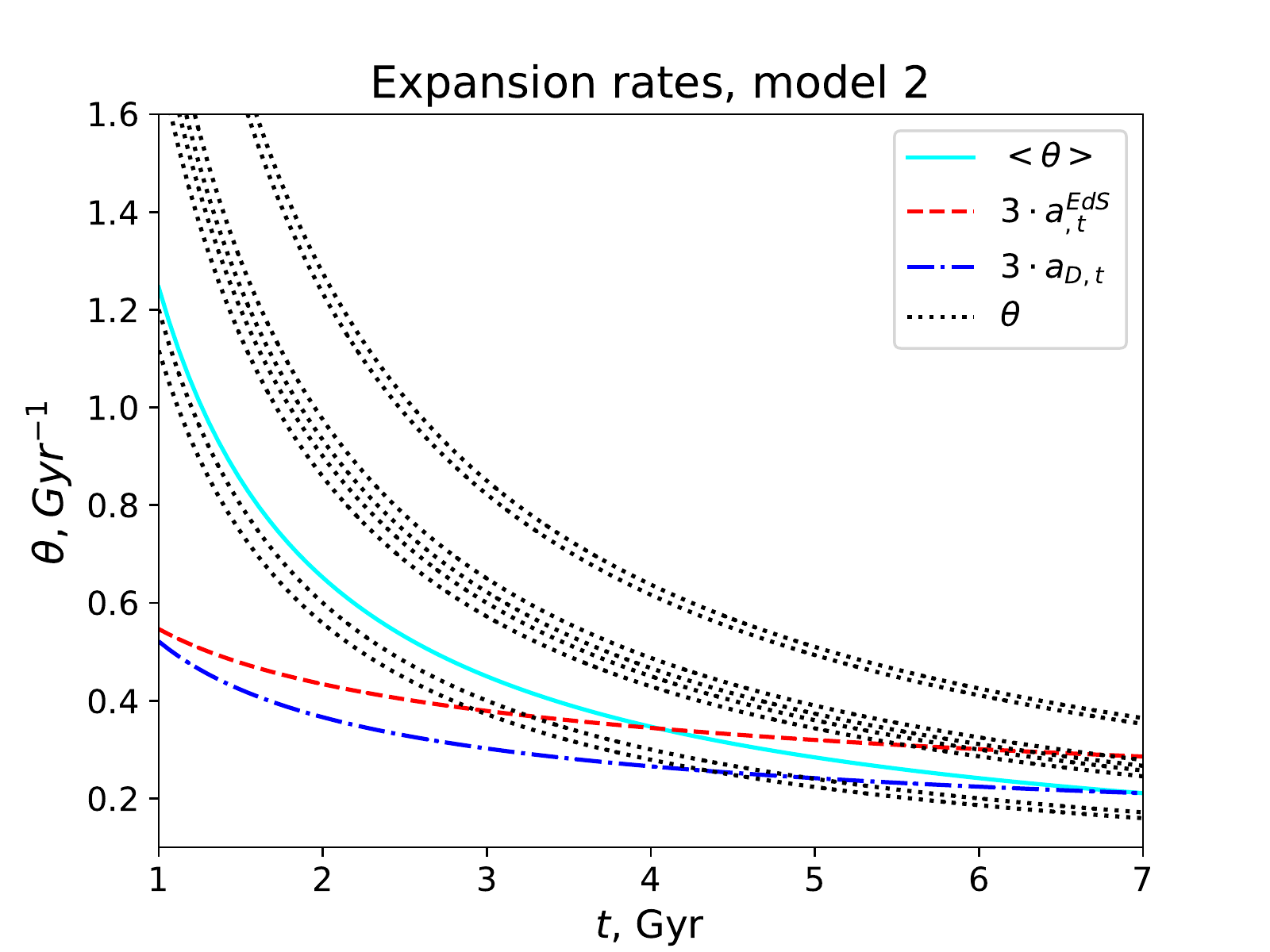}
	}
	\caption{Local expansion rates of each region of the fundamental blocks together with the average expansion rate of the models and the expansions of the EdS model (for comparison) and the average expansion $a_{D,t}$. For model 1, a close-up of $a_{D,t}$ is included to show that it is increasing with time.}
	\label{fig:expansion}
\end{figure*}

The studied model is of the type introduced in \citet{Hellaby}, where space is tessellated by cubes of a specific type of the homogeneous Bianchi I models (generalized Kasner models \citet{Kasner}), resulting in an inhomogeneous cosmological model.
\newline\indent
The considered local line element of a cube is
\begin{equation}
ds^2 = -dt^2 + \left(\frac{t}{t_0} \right) ^{2\alpha}dx^2 + \left(\frac{t}{t_0} \right) ^{2\beta}dy^2  + \left(\frac{t}{t_0} \right) ^{2\gamma}dz^2, 
\end{equation}
where $\alpha, \beta$ and $\gamma$ are constants and $t_0$ is present time.
\newline\newline
This metric fulfills the Einstein equation for a comoving perfect fluid with homogeneous density and homogeneous, anisotropic pressure ($p_x\neq p_y\neq p_z$ in general) - see \citet{Hellaby} for details. The local expansion rate of the fluid is $\theta = \frac{\alpha+\beta+\gamma}{t}$. The shear scalar is $\sigma^2=\frac{2}{3t^2}\left[ \left(\alpha^2+\beta^2+\gamma^2 \right)-\left(\alpha\beta + \alpha\gamma+\gamma\beta \right) \right]$.
\newline\indent
As in \citet{Hellaby}, eight cubes are arranged in a ``fundamental'' block that is used to tessellate all of space in order to construct an inhomogeneous cosmological model which is statistically homogeneous. To fulfill the Darmois junction conditions \citet{Darmois}, those of the metric parameters $\alpha, \beta$ and $\gamma$ which correspond to a direction orthogonal to a junction must be constant across the junction, while there is no restriction on the parameter in the direction parallel to the junction, i.e. for a junction with $x=$const., $\alpha$ may change across the junction while $\beta$ and $\gamma$ must be constant. The arrangement of the eight cubes is illustrated in figure \ref{fig:model} with numerical values for two particular models given in table \ref{table:models}.
\newline\indent
The parameter values given in table \ref{table:models} do not correspond to realistic values of density and pressure. Indeed, since $\rho\propto \frac{\alpha\beta+\beta\gamma+\alpha\gamma}{t^2}$ some regions have negative density, and some regions will not have a big bang singularity in all spatial directions. This is of no issue here as the models are not meant to be realistic renderings of the Universe: The purpose is to study the principles of light propagation in an inhomogeneous universe which does not contain an FLRW background, preferably with non-negligible backreaction. The principles of light propagation do not depend on particular values of e.g. pressure and density - not even the signs matter. Nonetheless, realistic values of $\rho$ and $p$ must generally be considered favorable. Such requirement was however found difficult to fulfill while also obtaining average accelerated expansion and non-negligible kinematical backreaction without introducing large regions expanding or contracting very fast in one or more directions, making the models very impractical for a light propagation study. The question of whether $\rho$ and $p$ take on realistic values was therefore not considered when choosing parameter values. The parameter values of model 1 were chosen to lead to a late-time average accelerated expansion without local accelerated volume expansion while keeping expansion rates and regions small enough for ``structures'' not to evolve much during the time it takes light rays to traverse the homogeneity scale (a natural requirement for expecting a simple relation between observables and spatial averages \citet{av_obs1,av_obs2}). The Bianchi I models have $^{(3)}R = 0$ so we trivially have $\Omega_R = 0$. Then $\Omega_Q= 1-\Omega_\rho$ which is of order $0.001$ at present time in the model and grows to order $0.01$ at times so early they are just barely traced by the light rays. Despite the negligible backreaction, the model is interesting for light propagation studies as the model everywhere locally behaves quite differently from its spatial average. This is seen in figure \ref{fig:expansion} which shows the expansion rates of both models. Model 2 does not have average accelerated expansion but it roughly has $\Omega_{Q}\in\left[0.04,0.1 \right] $ in the total time interval along the studied light rays. Note that the kinematical backreaction can be non-vanishing despite the average curvature being identically zero because the models have inhomogeneous pressure (see \citet{fluid2II}).

\begin{table*}
		\caption{Model parameters of Bianchi I cubes in the fundamental blocks. The parameters a and A, b and B, g and G refer to values of $\alpha$, $\beta$ and $\gamma$, respectively, in different cubes of the fundamental block as illustrated in figure \ref{fig:model}. Similarly, dx1, dx2 etc. refer to comoving side lengths of cubes of the fundamental block according to the illustration in figure \ref{fig:model}. }
	\centering
	\begin{tabular}{c c c c c c}
		\hline\hline
		Model & (a,b,g) & (A,B,G) & (dx1,dy1,dz1) & (dx2,dy2,dz2)& $t_0$ (Gyr) \\
		\hline
		1& ($1,1,1$) & $-0.05\cdot(1,1,-1)$ &  $30\cdot$(1,1,1) &$10\cdot$(1,1,1) &  8.9 \\
		2& $\left( \frac{4}{5},\frac{3}{4},1\right) $ &  $\left(\frac{1}{5},\frac{2}{3},\frac{1}{4} \right) $ & $10\cdot$(1,1,1) & $30\cdot$(1,1,1) & 7 \\
		\hline
	\end{tabular}

	\label{table:models}
\end{table*}

\section{Light propagation and mean observations}
\begin{figure*}
	\centering
	\subfigure[]{
		\includegraphics[scale = 0.5]{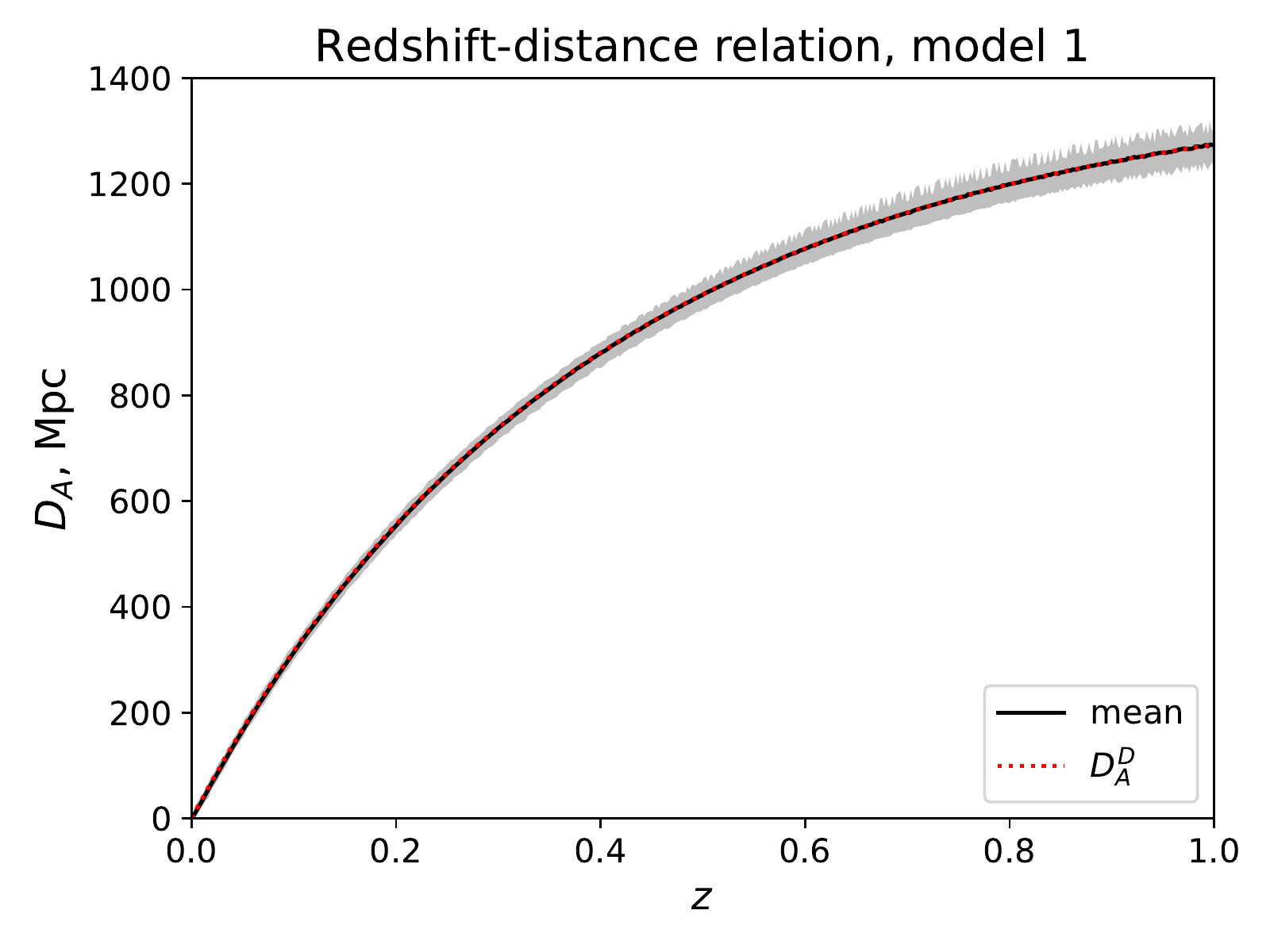}
	}
	\subfigure[]{
		\includegraphics[scale = 0.5]{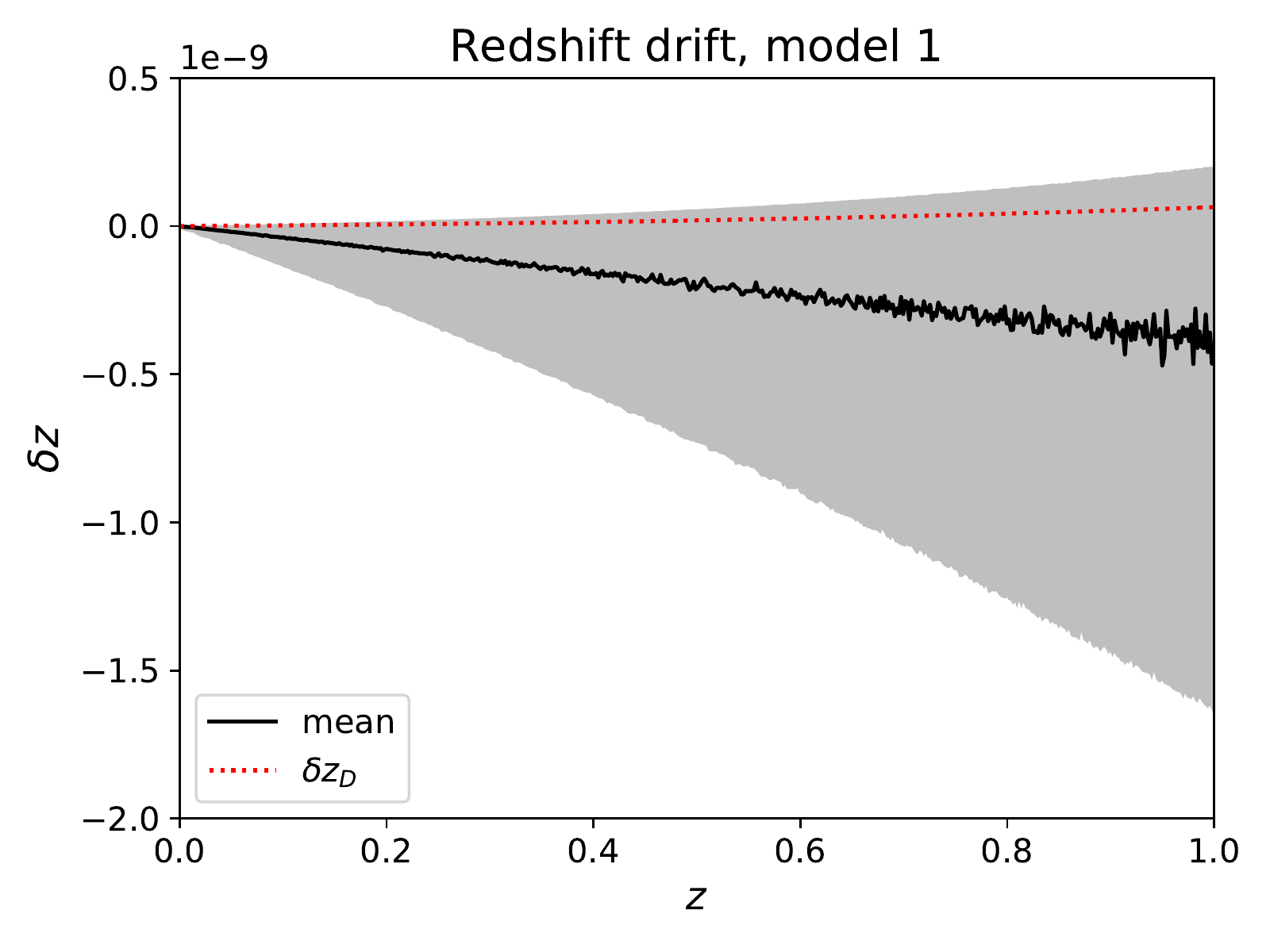}
	}
	\subfigure[]{
		\includegraphics[scale = 0.5]{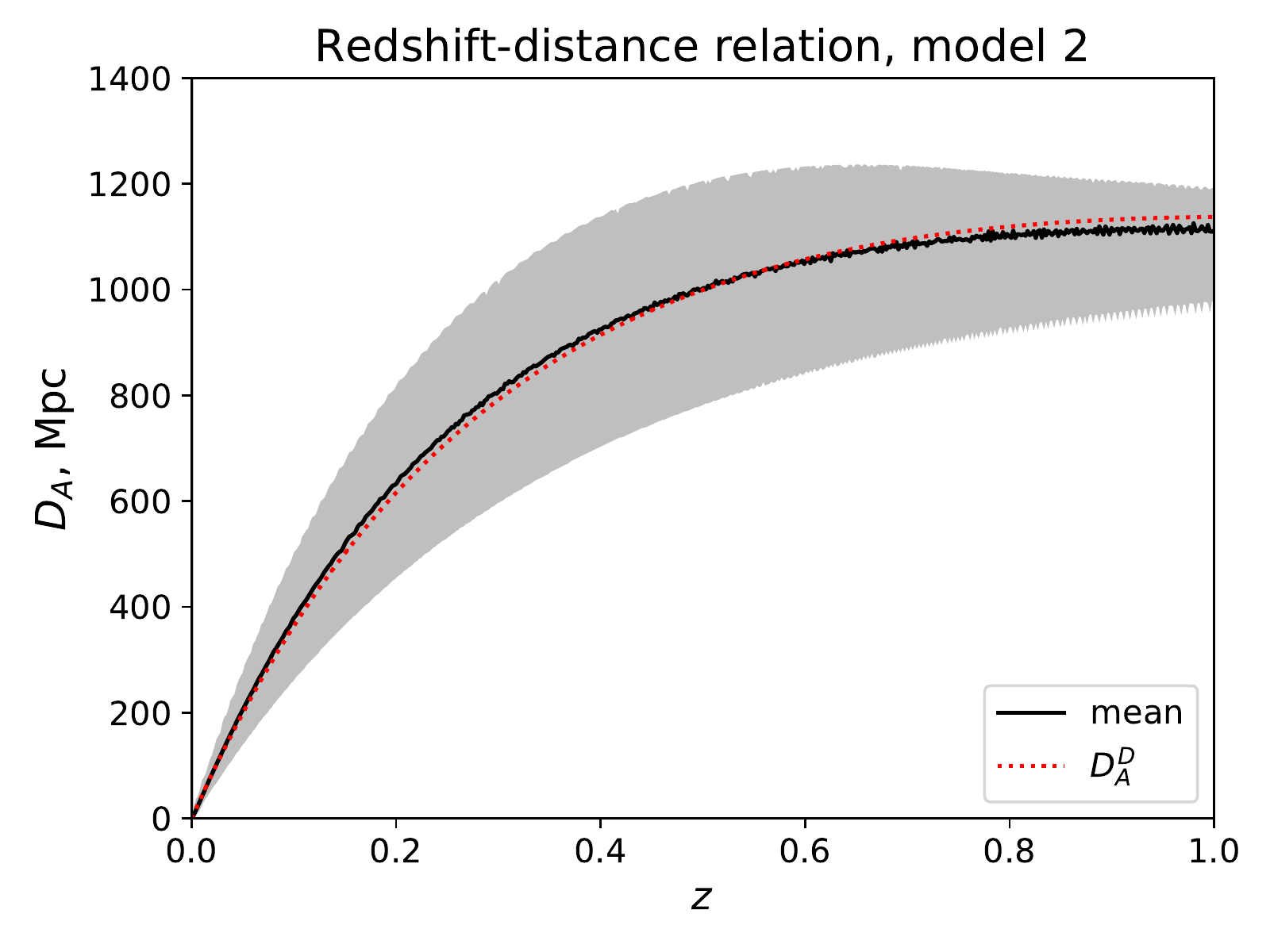}
	}
	\subfigure[]{
		\includegraphics[scale = 0.5]{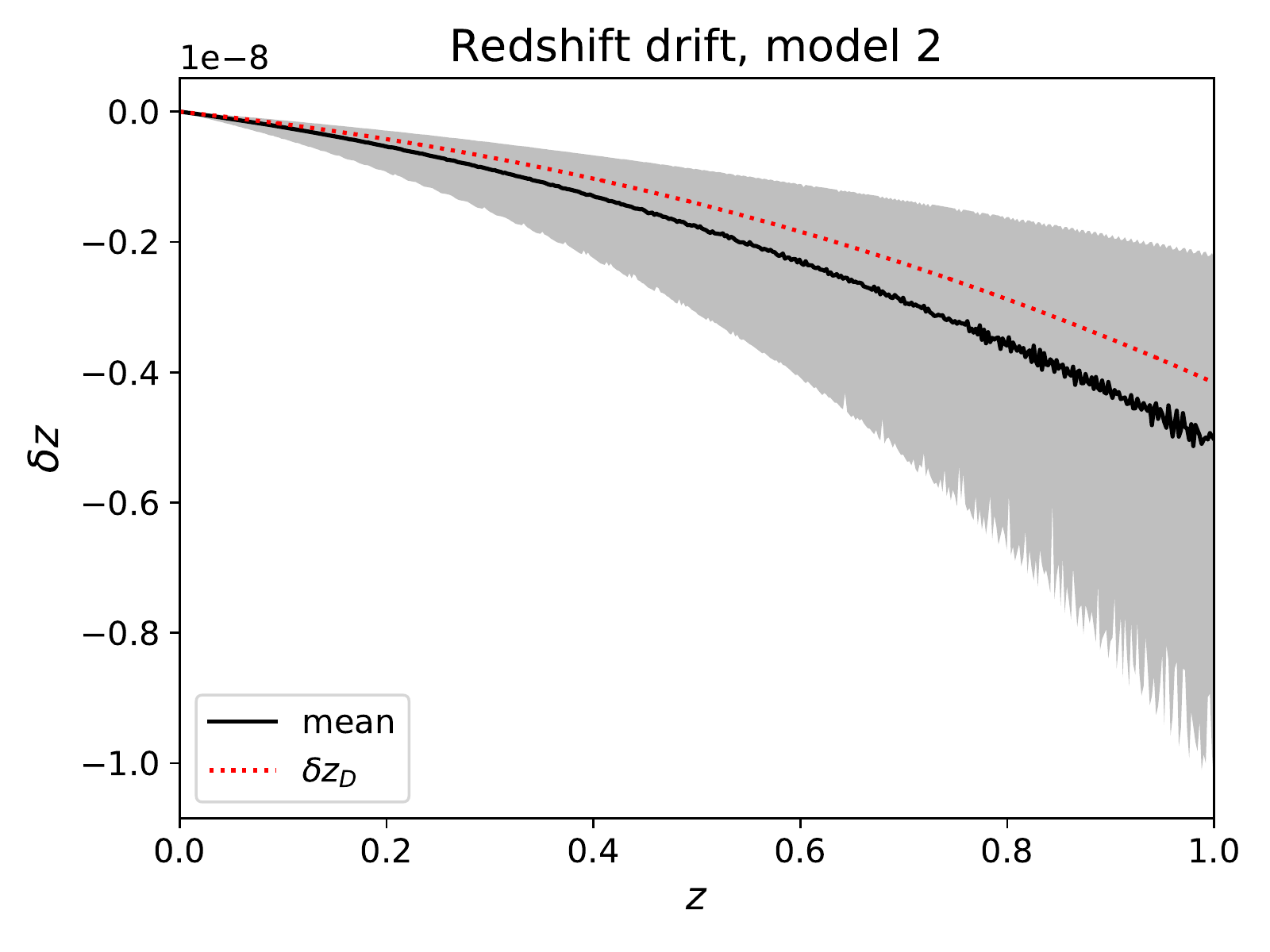}
	}
	\caption{Average, mean and spread of redshift-distance relation and redshift drift along 500 light rays compared to predictions based on spatial averages. The average and mean redshift-distance relations are indistinguishable in the figure for model 1. The redshift drift was computed using $\delta t_0 = 30$yr.}
	\label{fig:observables}
\end{figure*}
The exact redshift, angular diameter distance and redshift drift are computed along 500 light rays in each model. For each light ray, the spatial position of the present-time ($t=t_0$) observer and the  direction of observation are random. Since the model is inhomogeneous, the spatial positions of the observers must be random as the mean results would otherwise be biased according to the local Bianchi I model at the point of observation. The lines of sight must be random because the spacetime is not statistically isotropic. A lack of statistical isotropy is expected to impair any relationship between spatial averages and mean observations so it is important to remove the effect by observing in many, random directions. 
\newline\newline
Light paths are computed from the geodesic equations, $\frac{d}{d\lambda}\left(g_{\alpha\beta}k^{\beta} \right) = \frac{1}{2} g_{\mu\nu,\alpha} k^{\mu} k^{\nu} $ (the Einstein summation convention is used and Greek indices are spacetime indices while Latin indices are pure space indices). On the junctions between different Bianchi I cubes with the junction of the form $x^i$ = const., $g_{ii,i}$ contains a $\delta$ (delta-Dirac) function describing the change in $\alpha, \beta$ or $\gamma$ across the junction. For instance, at a boundary at constant $x=x_b$ where $\alpha$ changes by $\Delta\alpha$, the equation for $\frac{dk^x}{d\lambda}$ contains the term
\begin{equation}
-\frac{1}{2}\frac{g_{xx,x}}{g_{xx}}\left( k^x\right) ^2 = -\left( k^x\right) ^2\log\left( \frac{t}{t_0}\right) \delta(x-x_b)\Delta \alpha.
\end{equation}
Integrating this, one sees that this implies that $k^x$ is modified according to 
\begin{equation}
	k^x\rightarrow k^x\left( 1+\log\left( \frac{t}{t_0}\right) \Delta\alpha\right) 
\end{equation}
when the light ray crosses the boundary. Numerically this is seen to correspond to a renormalization of $k^x$ so that the light ray remains null. Thus, in general, the effect of the $\delta$ function is simply to re-normalize $k^i$ so that the geodesic remains null. 
The $\delta$ functions do not appear in the Riemann tensor and hence in the transport equation $
\frac{d^2D^a_b}{d\lambda^2} = T^a_c D^c_b$ from which the angular diameter distance is obtained (through $D_A = \sqrt{|\det D|}$). The tidal matrix has the components
\begin{equation}
T_{ab} = 
\begin{pmatrix} \mathbf{R}- Re(\mathbf{F}) & Im(\mathbf{F}) \\ Im(\mathbf{F}) & \mathbf{R}+ Re(\mathbf{F})  \end{pmatrix} ,
\end{equation}
where $\mathbf{R}: = -\frac{1}{2}R_{\mu\nu}k^{\mu}k^{\nu}$ and $\mathbf{F}:=-\frac{1}{2}R_{\alpha\beta\mu\nu}(\epsilon^*)^{\alpha}k^{\beta}(\epsilon^*)^{\mu}k^{\nu}$. $R_{\mu\nu}$ is the Ricci tensor, $R_{\alpha\beta\mu\nu}$ the Riemann tensor and $\epsilon^{\mu} := E_1^{\mu} + iE_2^{\mu}$ with  $E_1^{\mu}, E_2^{\mu}$ orthonormal vectors spanning the space orthogonal to the propagation of a light ray in the rest frame of the observer (here taken to be comoving with the fluid with $u^{\alpha} = \delta _0^{\alpha}$).
\newline\newline
The redshift drift, $\delta z$, describes the change/drift in the redshift of a comoving source as measured by a comoving observer as a function of observer proper time \citet{Sandage,McVittie} (non-comoving effects have also been studied \citet{peculiar_drift, general_observer} but are not considered here). In an FLRW universe, $\delta z = \delta t_0 (1+z)\left( a_{,t}(t_0) - a_{,t}(t_e)\right)$ which means that $\delta z$ measures spacetime expansion, and late time accelerated expansion will show as $\delta z>0$ for small $z$. For the model studied here, $\delta z$ can be computed according to
\begin{align}
\begin{split}
\frac{\delta z}{\delta t_0} &= \frac{dz}{dt_0} = \frac{d}{dt_0}\left( \frac{\left( k^{\mu}u_{\mu}\right) _e}{\left(k_{\alpha}u^{\alpha} \right) _0}\right)
\\ &= \frac{d}{dt_0}\left( \frac{k^t_e}{k^t_0}\right) = -\frac{k^t_e}{\left(k^t_0 \right) ^2}\frac{dk^t}{dt}|_0 + \frac{1}{k^t_0}\frac{dt_e}{dt_0}\frac{dk^t}{dt}|_e
\\ &= -\frac{1+z}{\left(k^t_0 \right) ^2}\left[ \frac{dk^t}{d\lambda} - k^ik^t_{,i}\right]|_0 + \frac{1}{k^t_0k^t_e}\frac{1}{1+z}\left[ \frac{dk^t}{d\lambda} - k^ik^t_{,i}\right]|_e,
\end{split}
\end{align}
where the subscript $e$ denotes evaluation at spacetime point of emission. The top line in the equation is valid for any spacetime so the above illustrates a fairly simple general method for computing the redshift drift. It was used that $\frac{dt_0}{dt_e} = 1+z$ and that $\frac{dk^t}{d\lambda}=k^{\mu}k^t_{,\mu}$. The last line is included to emphasize that $\delta z$, unlike $z$, is not given solely by an integral along the light ray, but also depends on local qualities of spacetime through boundary terms, i.e. through $\frac{dk^t}{d\lambda}$. For a homogeneous spacetime, $k^t_{,i} = 0$ and the above can be used to compute the redshift drift if one solves the geodesic equations. This is e.g. the case for FLRW and Bianchi I spacetimes, but for a spacetime with several Bianchi I regions joined, $\delta$ functions on boundaries between different regions lead to $k^t_{,i}\neq 0$. In order to obtain an expression for $k^{t}_{,i}$ (or $k^t_{,t}$), the geodesic equations are differentiated as in \citet{Ishak}, yielding $\frac{d}{d\lambda}k^{\mu}_{,\nu} = \frac{\partial }{\partial x^\nu}\frac{dk^{\mu}}{d\lambda}-k^{\beta}_{,\nu}k^{\mu}_{,\beta}$ which are solved simultaneously with the geodesic equations and the transport equation. Most of the equations for $k^{\mu}_{,\nu}$ contain $\delta$ functions related to the junctions between different regions. When crossing the boundary of $x^i = $const., the $\delta$ functions in $\frac{dk^t_{,i}}{d\lambda}$ and $\frac{dk^i_{,\mu}}{d\lambda}$ are non-zero. Their contributions are taken into account by re-normalizing $k^i_{,\mu}$ with the partial derivatives of the null condition and by re-normalizing $k^{t}_{,i}$ through the definition $\frac{dk^t}{d\lambda} = k^{\alpha}k^t_{,\alpha}$.
\newline\indent
Since each Bianchi I region is locally homogeneous, $k^{\mu}_{,i} = 0$ can be used as initial conditions when solving $\frac{dk^{\mu}_{,\nu}}{d\lambda}$. The initial conditions for $k^{\mu}_{,t}$ are then simply $\frac{1}{k^t}\frac{dk^{\mu}}{d\lambda}$.
\newline\newline
The results from applying the above set of equations to the models specified in table \ref{table:models} are shown in figure \ref{fig:observables}. Specifically, the figure shows the mean and spread of the redshift-distance relation and redshift drift along the 500 light rays in each model. These exact results are compared with the average redshift-distance relation, $\left(D_A^D, z_D \right) $, proposed in \citet{av_obs1,av_obs2} to be given by $H_D\frac{d}{dz_D}\left(\left(1+z_D \right) ^2H_D \frac{dD_A^D}{dz_D}\right) = -4\pi G\left\langle \rho+p\right\rangle D_A^D$, where $z_D:=\frac{1}{a_D}-1$ and the appropriate average of the anisotropic pressure is set as $\left\langle p \right\rangle := \frac{1}{3}\left\langle p_x + p_y + p_z \right\rangle $ (assessed by considering spatial directions of light rays along coordinate axes and comparing to \citet{av_obs2}). Note that a term representing the average null-shear has been dropped in the equation because the random directions of the sampled light rays should effectively lead this term to be negligible.
\newline\indent
Figure \ref{fig:observables} shows a good agreement between the mean and average redshift-distance relation in both models with the notable difference that in model 1, the agreement is nearly exact while it is clearly only approximate for model 2. In both models, there is a significant difference between the mean redshift drift and the drift of the mean redshift, $\delta z_D:=\delta t_0\left( \left(1+z_D \right)H_0^D - H_D  \right) $ (see e.g. \citet{koksbang3}). This result is in agreement with that found in \citet{koksbang2} which was, however, based on a toy-model of disjoint FLRW regions and not an exact solution to the Einstein equation. Note also that the spread around the mean of $\delta z$ is very large. This is presumably due to the anisotropy of the model and the large local effects which are not expected to be seen in the real universe.

\section{Summary}
By studying light propagation in two exact cosmological solutions to the Einstein equation it was shown that spatial averages can be used to describe the mean redshift-distance relation through an FLRW-like ``average'' redshift-distance relation. It was then shown that the drift of the average redshift entering this average redshift-distance relation is not equal to the mean redshift drift. This difference in the average-vs.-mean-relation for the two types of observations implies that redshift drift can be useful in relation to quantifying the importance of backreaction; the results found here indicate that a non-negligible backreaction should be expected to lead to a clear disagreement between observations based on the redshift-distance relation and redshift drift. It is especially interesting that the mean redshift drift is negative while the drift of the mean redshift is positive, in agreement with what was found in \citet{koksbang2}, but the anisotropy of the models make it unclear if the mean redshift drift will necessarily be negative if there is no local accelerated expansion.

\section{Acknowledgments}
The author thanks Syksy Rasanen for comments on the manuscript. Part of the numerical work was done using computer resources from the Finnish Grid and Cloud Infrastructure urn:nbn:fi:research-infras-2016072533.
\newline\newline
During the review process the author transitioned from being supported by the Independent Research Fund Denmark under grant number 7027-00019B to being supported by the Carlsberg Foundation.

\section{Data Availability Statement}
The data and/or code used to generate the data and results presented here will be shared on reasonable request to the author.





\bsp	
\label{lastpage}
\end{document}